\documentclass[12pt, letterpaper]{article}

\usepackage{amsmath,amssymb}
\usepackage{cite}
\usepackage{fancyhdr}
\usepackage[top=1in, bottom=1in, left=1in, right=1in]{geometry}
\usepackage{graphicx}
\usepackage{hyperref}

\numberwithin{equation}{section}
\date{}

\title{{\rm\footnotesize \qquad \qquad \qquad \qquad \qquad \ \qquad \qquad \qquad \ \ \ \ \ \                       RUNHETC-2015-07     
SCIPP 15/11}\vskip.5in    Density Functional Theory for Field Theorists {\bf I}}
\author{Tom Banks\\
NHETC and Department of Physics \\
Rutgers University, Piscataway, NJ 08854-8019\\
{\it and}\\
Department of Physics and SCIPP\\
University of California, Santa Cruz, CA 95064\\
E-mail: \href{mailto:banks@physics.rutgers.edu}{banks@physics.rutgers.edu}
}
\begin{document}

\maketitle
\thispagestyle{fancy} 

\begin{abstract} I summarize Density Functional Theory (DFT) in a language familiar to quantum field theorists, and introduce several apparently novel ideas for constructing {\it systematic} approximations for the density functional.  I also note that, at least within the large $K$ approximation ($K$ is the number of electron spin components), it is easier to compute the quantum effective action of the Coulomb photon field, which is related to the density functional by algebraic manipulations in momentum space.

\normalsize \noindent  \end{abstract}


\newpage
\tableofcontents
\vspace{1cm}


\section{Density Functional Theory for Quantum Field Theorists}

Much of atomic, molecular and (quantum) condensed matter physics, reduces, in the non-relativistic limit, to the problem of solving the Schr\"{o}dinger equation for point-like nuclei and electrons, interacting via the Coulomb potential.  The electrons are fermions, and are best treated by introducing an electron field \begin{equation} \psi_i ({\bf x}) = \sum_{\bf k} a_i ({\bf k}) e^{i {\bf k\cdot x}} , \end{equation} where $i$ is a spin label.
If we introduce dimensionless space-time coordinates, by using the Bohr radius to measure length, and the Rydberg to measure energy, the only parameters in the problem are the nuclear charges $Z_a$ and the mass ratios $\frac{m_e}{m_a}$.  The latter appear multiplying the nuclear kinetic terms
\begin{equation} H_{kin-nuc} = \sum_a \frac{m_e}{m_a} {\bf P}_a^2 .\end{equation}

The mass ratios are all $\leq .0005$, and this leads to the Born-Oppenheimer approximation for calculating a systematic series in powers of these small dimensionless parameters.   
One first freezes the nuclear positions and finds the electronic ground state energy $V_{BO} ({\bf R}_a )$ as a function of the nuclear positions.  Then one minimizes the Born Oppenheimer potential and expands around the minimum to quadratic order.

The minimum value of $V_{BO}$ is the order $1$ contribution to the ground state energy of the system in Rydbergs.  The oscillator corrections are down by $\sqrt{\frac{m_e}{m_a}}$ .  

The next stage depends on whether one is dealing with atoms, molecules or large crystals.   For atoms, the ground state is spherically symmetric.  For molecules, the minimum of the Born Oppenheimer potential has a shape (the one you find models of in chemistry departments), and can be rotationally excited.
The rotational level splittings are of order $\frac{m_e}{m_a}$.
For large crystals the coherent superpositions of rotational levels corresponding to fixed positions of the crystal, satisfy almost classical equations of motion.   On the other hand, the low energy spectrum of vibrational levels reaches down to
$\sqrt{m_e / m_a} /L $, where $L$ is the linear size of the crystal.  These phonon excitations are important to a number of properties of solids.

The hard quantum mechanical part of all of these problems is the solution of the electronic Schr\"{o}dinger equation for fixed nuclear positions.  It is a many fermion problem with no dimensionless parameters.  For infinite crystals one can introduce the electron density in Bohr units as a dimensionless parameter.   Density Functional Theory is a general approach to this difficult problem.   It was invented by Kohn Hohenberg and Shams\cite{KHS}.

In this section, I want to present an elementary approach to DFT using concepts familiar to particle theorists, in the hope that we can make some contributions that may have been missed in the condensed matter literature.  DFT has had remarkable successes in all areas of atomic, molecular and condensed matter physics, but much of the work is numerical.  My hope is that there are some more analytical approaches to the calculation of the density functional, which can be of some use.
In the second part of this paper I will present an analytic approach, based on an expansion in the number of electron spin components.  

The Born Oppenheimer approximation reduces all of our problems to that of electrons interacting via Coulomb repulsion, in a background potential $V_{ext} $, which neutralizes their total charge.  $V_{ext}$ is a finite sum of nuclear Coulomb potentials, but it is convenient to generalize it to be an arbitrary function of ${\bf x}$ .   Our problem then is to find the ground state energy of interacting electrons in a general neutralizing external potential, as a functional of the potential.

Field theorists will instantly recognize this as the problem of finding the generating functional $W[V_{eff} ({\bf x}) ]$ of connected equal time Green's functions of the electron density operator
\begin{equation} N( {\bf x} ) \equiv \psi^{\dagger}_i ({\bf x}) \psi_i ({\bf x} ) ,\end{equation}
in the Homogeneous Electron Gas (HEG).  This observation is, from one point of view, the foundation of density functional theory.   The density functional is just the functional Legendre transform of $W$.

More intuitively, we can use the variational principle to find the ground state energy by minimizing
\begin{equation} \langle \chi | \int d^3 x\ \psi^{\dagger}_i ({\bf x}) ( - \nabla^2 -\mu ) \psi_i ({\bf x}) + \frac{1}{2} \int d^3 x\ d^3 y\ \frac{(\tilde{N}({\bf x})\ \tilde{N}({\bf y})}{|{\bf x - y}|}  +\int d^3 x\  N({\bf x}) V_{ext} ({\bf x}) , | \chi \rangle , \end{equation}
over all normalized states $| \chi \rangle$.  $\mu$ is the chemical potential and $\tilde{N} ({\bf x}) = N ({\bf x}) - n_0 ,$ where $n_0$ is the neutralizing charge density.  We perform the minimization in stages, first minimizing the expectation value of the Hamiltonian among normalized states with a fixed value of the density
\begin{equation} n({\bf x}) = \langle \chi | N(x) | \chi \rangle , \end{equation} and then minimizing the resulting functional of $n({\bf x})$ w.r.t. the density.

This procedure recommends itself, because the non-universal part of the energy, the part that depends on the external potential, is just 
\begin{equation} \int d^3 x\  n({\bf x}) V_{ext} ({\bf x}) , \end{equation} and is thus a simple and explicit functional of the density alone.  The hard part of the calculation, which gives rise to the density functional $F[n({\bf x})]$ is a property of the HEG: it is the minimum expectation value of the HEG Hamiltonian in states that have fixed expectation value of the density operator $N({\bf x})$.  

The message is clear, if surprising.  The better we can do at solving the apparently academic problem of the HEG, the better we can do at {\it all} the problems of atomic, molecular, and condensed matter physics.  Since the ground-breaking Hohenberg-Kohn paper that pointed this out\cite{hk} condensed matter physicists and quantum chemists have been making this insight pay off in a major way.  The HEG itself is a difficult problem, with no intrinsic dimensionless expansion parameter besides the average electron density in Bohr units.  For high density, an expansion around the free electron gas is useful, though it must be re-summed to deal with apparent IR divergences.   The point is that the long distance Coulomb potential is screened (the quantum analog of Debye screening in a classical charged gas at finite temperature) so the long distance and weak coupling limits do not commute.  This problem is easily dealt with and several terms in the high density expansion of the ultra-local part of $F[n]$ are known.

Much of the work on DFT uses an approximation first proposed by Kohn and Shams (KS)\cite{ks}.  They write
\begin{equation} F[n] = F_0 [n] + \frac{1}{2} \int d^3 x\ d^3 y\ \frac{(\tilde{n}({\bf x})\ \tilde{n}({\bf y})}{|{\bf x - y}|}  + F_{LDA} [n] , \end{equation} where the first term is the density functional of a free electron gas\footnote{There is a subtlety in this statement, in that Kohn and Shams write $F_0 [n]$ in terms of single particle solutions to the Schr\"{o}dinger equation $ - \nabla^2 + \mu - \frac{\delta F_0 [n]}{\delta n({\bf x})} = \epsilon_i \psi_i $, rather than the density itself.  We'll deal with this subtlety in the section on the KS equations.}, the second, called the Hartree term, is the interaction term of the Hamiltonian with the density operator replaced by its expectation value, and the third, which is usually calculated by sophisticated numerical algorithms\cite{numerical}, is an ultra-local functional of the density.  It is the ground state energy density of the HEG. A variety of fitting functions have been proposed, which fit the numerical data and match the high and low density expansions\cite{fit}, but there does not seem to be a systematic approach to computing corrections.  There have been various attempts to go beyond the local density approximation. Some particularly successful ones are found in \cite{gga}, but the most accurate fit to real data is found by making empirical fits to parameters in theoretically motivated formulae\cite{becke}. There is, to my knowledge, no systematic expansion for the density functional.  One aim of the present paper is to propose directions for such a systematic set of approximations. We will see that the leading term in an expansion in the inverse number, $K$, of electron spin components gives something very close to the first two terms in the KS ansatz.  The only difference is that $F_0 [n]$ is expressed in terms of the density, rather than KS orbitals.  However, at the density which gives minimum energy, the two approaches give the same result. The higher order terms in the $1/K$ expansion are all describable in terms of diagrams whose vertices are functional derivatives of the partition function of a non-interacting electron gas w.r.t. a time dependent background potential. The propagator is similarly related to the density two point function.  These remarks suggest an important step forward will be to develop analytic functional approximations to the non-interacting problems with an external field.  I will make some tentative remarks about this in the penultimate section of this paper.

The large $K$ expansion may also make it possible to explore analytically the properties of low density states of the HEG, which violate translation symmetry spontaneously, like the Wigner crystal\cite{wigner}, and emulsions or stripes\cite{kivelson}.  The leading order large $K$ expansion gives a steepest descent approximation to the expectation value of the potential coupled to the density operator.  This is a translation invariant equation, but can have solutions which violate translation invariance and minimize the energy.  Large $K$ expansions are not always reliable guides to the detailed physics of such phases and the transitions between them, but they certainly provide a convenient starting point for quantitative discussion. We'll see that in leading order, the large $K$ expansion predicts a first order transition between a Wigner crystal and a homogeneous phase, at zero temperature.

The KS approximation certainly gets some qualitative features of molecules and solids right.   If we combine the Hartree term, the external potential term, and the Coulomb repulsion between nuclei, and ignore everything else, then the energy is minimized by making the electron density exactly cancel the nuclear charge density.  However, because of the uncertainty principle, $F_0$ is large if the density varies on scales much shorter than the Bohr radius, while the nuclear wave functions are localized on scales shorter by a factor of $\sqrt{m_e / m_a}$, so exact cancelation is impossible.  However, the Bohr radius of a single electron ion with a nucleus of charge $Z$ is $1/ Z$ in Bohr units, so the electron density tends to cluster in the vicinity of nuclear positions and to try to screen out the large nuclear charges, over distances of order $1$ in Bohr radii.  Once the nuclear charges are screened down to something of order $1$, the competition between attraction to the nuclei, mutual repulsion, and Fermi statistics, becomes less easy to estimate, and one must go to more refined approximations.  However, it's clear that the qualitative fact that only a few valence electrons per nucleus are responsible for molecular binding, and transport in solids, follows simply from this crude approximation.

What's especially remarkable is that this essentially classical minimization problem captures so much of the quantum mechanics of a many electron system.  Field theorists are used to the platitude that, in principle, everything one wants to calculate can be ``reduced" to the problem of finding stationary points of the quantum effective action, but in atomic, molecular, and condensed matter physics, these fantastical claims actually lead to manageable computations.

To conclude this introduction, I want to address myself to quantum chemists, band structure theorists, or other professional utilizers of DFT, who might happen across it.  This paper is not addressed to you.  As a practical tool, the methods proposed in this paper are not ready for prime time.  My main motivation in writing has been to get professional field theorists interested in the field theory of the HEG, which is enormously simpler than the relativistic theories we typically study, but of enormous practical utility.  The professionals have developed DFT into a tool of impressive accuracy, but the foundations of the theory and a more systematic approach to it have not been a center of interest for a while, as far as I am aware.  A. Zangwill has pointed out a very interesting early paper\cite{dietz}, which takes a field theoretic approach to DFT, similar to that advocated here.  It will be interesting to compare the approach taken there with my own.  The application of modern techniques of QFT to the HEG, is certainly a great mathematical physics project, and it might lead to real advances in practical calculations.  I'm taking my first tentative steps in this field, and I would certainly appreciate guidance to the existing literature and things I might have missed.

Some textbooks/monographs on DFT can be found in \cite{books}, and an excellent recent history of the subject, from which I've benefitted enormously is \cite{zangwill}.

\section{Expansions in the Number of Spin Components}
\subsection{The Large $K$ Expansion}

From the point of view of Quantum Electrodynamics, the Hamiltonian of the HEG results from integrating out the electro-magnetic field in the non-relativistic approximation for electron motion.  We can re-introduce the scalar potential by a Hubbard-Stratonovich transformation\footnote{This is done all over the condensed matter literature.  I've followed some concise lecture notes by Ben Simons\cite{simon}, with small changes in notation.}.  The finite temperature free energy of the HEG is then written as a Euclidean functional integral, with action for the Coulomb photon field $\phi$:

\begin{equation} S = \int d\tau\ \sum_{\bf k} {\bf k}^2 \phi_{\bf k} \phi_{\bf -k} + 2 {\rm Tr\ ln\ } \bigl(\partial_{\tau}  - \nabla^2 - i g \phi (\tau, {\bf x}) - V_{ext} ({\bf x})\bigr) .\end{equation} The trace is taken over the space of  fermionic wave functions, $\psi (\tau, {\bf k})$ , periodic on a spatial torus and anti-periodic in Euclidean time. We've introduced the torus in order to isolate the spatial zero mode of the Coulomb photon field $\phi$. This mode is omitted from the sum, corresponding to the fact the the zero mode of the nucleon charge density, exactly cancels that of the electron charge density for a neutral system.  We absorb the chemical potential into $V_{ext}$. The dimensionless parameter $g$ is equal to $1$ when we work in Bohr-Rydberg units.  We insert it in order to obtain a smooth large $K$ limit when we give the fermions $K$ spin components.

The log of the fermion determinant, like most functional determinants, has an additive infinity.  We normalize it by insisting that for $\tau$ independent $\phi$ it just equals the free energy of electrons in the external potential $V - i\phi$.  

We now modify the HEG by allowing the electrons to have $K$ spin components, turning the $SU(2)$ spin rotation symmetry into $SU(K)$.  The density operator $N(x) = \psi^{i\ \dagger} (x) \psi_i (x) $ has an expectation value of order $K$ and connected two point correlations of order $K$, so, in order to have a smooth large $K$ limit, we set 
$g = \sqrt{2/K} $, and define $\phi = \sqrt{K/2} \sigma$.  The resulting functional integral over $\sigma$ has an action

\begin{equation} S[\sigma] = K[ \frac{1}{2} \sum_{\bf k}\ {\bf k}^2 \sigma_{\bf k} \sigma_{\bf - k} + {\rm Tr\ ln\ } (\partial_{\tau}   - \nabla^2 -  i (\sigma - iV_{ext}) ] \equiv K\Bigl[ \frac{1}{2} \int\ \sigma (- \nabla^2) \sigma + \beta L[\sigma - iV_{ext}]\Bigr] .  \end{equation}

For large $K$ the fluctuations of $\sigma$ around the stationary point of the action are of size $\frac{1}{\sqrt{K}}$.  For an equilibrium state, the stationary point should be independent of Euclidean time.  Shifting variables to $ \chi \equiv i\sigma + V_{ext} $, the equation for a stationary point becomes
\begin{equation} \nabla^2 (\chi_0 - V_{ext}) = - \frac{\delta L}{\delta \chi ({\bf x})} . \end{equation}
The functional derivative of $L$ on the RHS is the electron density in a thermal equilibrium state of non-interacting electrons with external potential $\chi_0 $.  Note that in terms of the original variable $\sigma$, the stationary point is on the imaginary axis, corresponding to a real potential.  This often happens when evaluating an integral by the method of steepest descents.

The value of the free energy $E$ (defined by $Z = e^{- \beta E}$) at the stationary point is
\begin{equation} E [V_{ext}]  = K \int d^3x\ d^3y\ \Bigl[ \frac{\delta L}{\delta\chi ({\bf x})} \frac{1}{ - 2\nabla^2} ({\bf x, y })\frac{\delta L}{\delta\chi ({\bf y})} +  L[\chi_0 ] \Bigr] ,\end{equation}  where $\beta$ is the inverse temperature. The functionals are evaluated at the stationary point, at real $\chi_0$, which is independent of Euclidean time. 

The density $n({\bf x})$ is defined as the functional derivative of $E$ w.r.t. $U$.  To evaluate it we use the equation for the stationary point to write
\begin{equation} \frac{\delta U({\bf x})}{\delta \chi_0 ({\bf y})} = \delta^3 ({\bf x - y}) + \int\ G(y, w) \frac{\delta^2 L}{\delta \chi_0 ({\bf w}) \delta \chi_0 ({\bf x})} . \end{equation}  We then calculate 
\begin{equation} \frac{\delta E}{U({\bf x})} = \int\ d^3 y\ \frac{\delta E}{\delta \chi_0 ({\bf y})}\frac{\delta \chi_0 ({\bf y})}{\delta U({\bf x})} , \end{equation} using the chain rule, and obtain
\begin{equation} \frac{\delta E} {\delta U({\bf x})} = K \frac{\delta L}{\delta \chi_0 ({\bf x})} . \end{equation}

The free energy can now be written
\begin{equation} E[V_{ext}]  = \int d^3x\ d^3y\ \Bigl[ n({\bf x})
\frac{1}{ - 2K \nabla^2 } ({\bf x, y })  n({\bf y}) + K L[\chi_0 ] \Bigr] ,\end{equation}
where 
$$ n({\bf x}) = K \frac{\delta L}{\delta \chi_0 ({\bf x})} . $$  Since $K L[\chi_0 ]$ is just the free energy of the system of non-interacting electrons in the external potential $\chi_0$ we can write
\begin{equation}  K L[\chi_0 ] = F_0 [n] + \int\ [n \chi_0 ] . \end{equation}
Now use
\begin{equation} \nabla^2 (\chi_0 - V_{ext}) = - \frac{\delta L}{\delta \chi ({\bf x})} . \end{equation}
We see that the Legendre transform of $E[V_{ext}] $, which defines the leading large $K$ contribution to the density functional $F[n]$ is just given by

\begin{equation} F[n] = F_0 [n] + \frac{1}{2K} \int\ d^3 x\ d^3 y\ n({\bf x}) \frac{1}{\nabla^2} ({\bf x, y}) n({\bf y}) , \end{equation}
which is the standard Hartree approximation to the density functional if $K = 2$.
Note that since $n = 0(K)$, both of these terms are of order K in the large $K$ expansion.

The term of order $1$ in the large $K$ expansion of the free energy is \begin{equation} F_1 = \frac{1}{2} {\rm Tr\ ln\ } [ - \nabla^2 \delta^3 ({\bf x - y})\delta (t_x - t_y) + \frac{\delta^2 L}{\delta\chi ({\bf x}, t_x) \delta\chi ({\bf y}, t_y)}]  . \end{equation}  The operator whose determinant is evaluated here is called the inverse propagator of the $\chi$ field. 

Higher order terms in the large $K$ expansion involve Feynman diagrams whose vertices come from higher functional derivatives of $L$, and propagators for the $\chi$ field, which are inverses of the above integro-differential operator.  The key to evaluating these corrections is the solution of the single electron problem in an external potential, as a functional of the potential.

It's also interesting to understand the relationship of the density functional to the functional $\Gamma [ \chi ]$, which field theorists call the effective action or 1PI (one particle irreducible) generating functional of the field $\chi$.  This is defined  in the HEG with no external potential. The diagrams of the $1/K$ expansion have a propagator that is given, to zeroth order by $ [- \nabla^2 + \Pi (x - x^{\prime})]^{-1}$, where I've introduced space-time notation to save typing ($x^0 = t$) and $\Pi$ is the one loop, Euclidean time dependent two point function of the density operator.  One can divide all connected Feynman diagrams into those which cannot be cut in two, by cutting a single propagator line (the one photon irreducible diagrams), and those which can.  It's well known to field theorists that one can easily (meaning using only algebra in momentum space, without doing integrals) calculate the exact answer for any $n$ - point correlation of the field $\chi$, if one knows the exact $1PI$ correlators of $\chi $ for $k \leq n$.  For the two point function, the result is known as the Dyson series. One simply corrects, order by order in the $1/K$ expansion, the polarization function $\Pi (x - x^{\prime})$, defined as the sum of all 1PI two point functions, and writes the full propagator as $( - \nabla^2 + \Pi )^{-1}$.  

1PI n-point functions, $\Gamma_n ( x_1 \ldots x_n)$ of the field $\chi$ are written in terms of Feynman graphs involving loops of one component free electrons, which define the $k$ point vertices of the graphs as the connected, time dependent, $k$ point correlators of a gas of free electrons with a single spin component.  The $k$ point vertex has a weight $K^{- (k - 2)/2}$ .   The propagator for the photon contains only a single fermion loop contribution (RPA), and we sum over all 1PI graphs.  The generating functional of 1PI graphs is
\begin{equation} \Gamma[ \chi_c ] = \sum_{n=0}^{\infty} \frac{1}{n!} \int d^4n x\ \Gamma_n ( x_1 \ldots x_n) \chi_c (x_1) \ldots \chi_c (x_n )  . \end{equation}   
It is well known to quantum field theorists that
\begin{equation} \Gamma[ \chi_c ]  + \int d^4 x J(x) \chi_c (x) , \end{equation} evaluated at the value of $\chi_c$ satisfying \begin{equation} \frac{\delta\Gamma[ \chi_c ] }{\delta \chi_c (x)} = - J(x) \end{equation} is minus the logarithm of the functional integral over $\chi$ with action
\begin{equation} S[\chi ] = K [ \frac{1}{2} \int \chi \nabla^2 \chi + L[\chi] + \int \chi (x) J(x) . \end{equation} The equation above says that $\Gamma [\chi_c]$ is the Legendre transform of the negative log of this functional integral.  Call the Legendre transform of $\Gamma$, $W[J]$.  

On the other hand, the density functional (generalized to time dependent density fields) is the Legendre transform, w.r.t. $V_{ext}$ of the negative log of the functional integral with action
\begin{equation} S[\chi ] = K [ \frac{1}{2} \int (\chi - V_{ext}) \nabla^2 (\chi - V_{ext}) + L[\chi]  . \end{equation}
Thus $F[n]$ is the Legendre transform of \begin{equation} E[V] = W[ - \nabla^2 V_{ext} ] + \frac{1}{2} \int (V_{ext}) \nabla^2 (V_{ext}) . \end{equation}   The Legendre transform is not a linear operation.  However, the Legendre transformed $n$ point functions, are given by tree diagrams constructed from the $n$ point functions of the original functional (as vertices) connected by the inverse two point function.  The functionals $E$ and $W$ have $n > 3$ point functions which are the same up to multiplication by powers of momenta of the external legs, and two point functions which differ by a shift by $k^{-2} $ .It's easy to construct the momentum space $n$ point functions of $F[n]$ as algebraic combinations of the 1PI $n$ point functions of $\Gamma$, thus exploiting the diagrammatic simplification of one photon irreducibility.  

An even easier procedure is simply to evaluate $\Gamma [\chi_c ]$ and use the relation
\begin{equation}  E[V] = W[ - \nabla^2 V_{ext} ] + \frac{1}{2} \int (V_{ext}) \nabla^2 (V_{ext}) , \end{equation}  and the Legendre transform relation between $W$ and $\Gamma$ to evaluate $E[V]$ directly, as the solution to a variational problem involving $\Gamma$.  $F[n]$ has the virtue of being manifestly gauge invariant under space independent, but time dependent, gauge transformations of $\chi$, while $\Gamma$ is not.  However, the simplicity of this calculational scheme suggests that it is the right way to proceed.  

\subsection{Small K Expansion}

We can also contemplate a small $K$ expansion of the functional integral over $\sigma$.  For this purpose it is convenient to go back to the original variable $\phi$, in terms of which we see that the potential term in the Schr\"{o}dinger operator becomes very large in the small $K$ limit, while the whole determinant contribution becomes small.  We know that neglecting the determinant entirely will not work for small wave numbers, where screening effects are important.  This suggests that a good approximation to the determinant for large, slowly varying potentials will be useful in this limit.  This is precisely where the WKB approximation to the fermion determinant should work.  

If indeed we succeed in constructing a small $K$ expansion as well as a large $K$ expansion, then the technique of ``two point interpolation" (a generalization of two point Pade approximants to expansions that are not pure integer power laws) might be expected to give reliable results for all values of $K$.

\subsection{The KS Equations}

All of the applications of DFT are based on a series of approximations to $F[n]$ introduced by Kohn and Sham (KS)\cite{ks} .  For non-interacting fermions, the density functional is computed by introducing independent single particle orbitals $\psi_i ({\bf x}) $, in terms of which the density is \begin{equation} n({\bf x} ) = \sum_i \psi_i^* ({\bf x}) \psi^0_i ({\bf x}) . \end{equation}   These are the lowest $N$ normalized eigenstates of the single particle Schr\"{o}dinger equation
\begin{equation} - \nabla^2 \psi_i + \mu + V({\bf x}) \psi^0_i = \epsilon_i \psi^0_i . \end{equation}  The value of $N$ is taken to coincide with $\int n({\bf x}) $, which is tuned by tuning the chemical potential $\mu$.  The energy is 
\begin{equation} \sum_ i \epsilon_i = \int\ \sum_i (\psi^0_i )^* ({\bf x}) [- \nabla^2 + \mu + V({\bf x}) ] \psi^0_i ({\bf x}) = F_0 [n ({\bf x})]  + \int\ V({\bf x}) n({\bf x})  . \end{equation}  From this we see that {\it for the non-interacting problem}
\begin{equation} F_0 [ n ] =  \int\ \sum_i (\psi^0_i )^* ({\bf x}) [- \nabla^2 + \mu ]\psi^0_i ({\bf x}) . \end{equation}
Note that these $\psi^0_i $ are fixed functions, because the potential in the Schr\"{o}dinger equation, which they satisfy, is determined by
\begin{equation} \frac{\delta F_0 [n] }{\delta n({\bf x})} = - V({\bf x}) . \end{equation}  $F_0 [n]$ is then simply the kinetic energy of these single particle orbitals, and the density is written $ n({\bf x}) = \sum_i (\psi^0_i )^* ({\bf x}) \psi^0_i ({\bf x})$.

KS now take the single particle orbitals as their variational parameters, rather than the density.  The density functional is taken to be
\begin{equation} F[n] = K[\psi_i^* , \psi_i] + F_{Hartree} [n] + F_{xc} [n] , \end{equation} 
where $ n({\bf x}) = \sum_i \psi_i^* ({\bf x}) \psi_i ({\bf x})$, and $K$ the kinetic energy functional of the orbitals.  One then varies $ F[n] + \int V_{BO} n$  w.r.t. the $\psi_i^* $ after making a suitable approximation to $F_{xc} $ , obtaining the KS equations:
\begin{equation} (- \nabla^2 + \mu ) \psi_i  - [\frac{\delta (F_{Hartree} [n] + F_{xc} [n])}{\delta n({\bf x})} - V_{BO} ]\psi_i = \epsilon_i \psi_i .\end{equation}  The $\epsilon_i$ are Lagrange multipliers, which enforce the condition that the orbitals are normalized on the variational problem.  

These equations are solved by iteration.  One starts with the solutions without the Hartree and {\it X-C} contributions to the density functional, computes the density, and plugs it into the density functional to get a new estimate of the self consistent potential, then repeats the exercise.  My understanding is that the numerical convergence of the iteration is fast and stable.

When $\psi_i$ satisfy the KS equations, the functional $K [\psi_i* , \psi_i]$ is {\it not} equal to the non-interacting density functional $F_0 [n] $, because the potential in the Schr\"{o}odinger equation satisfied by the $\psi_i $ is not equal to $- \frac{\delta F_0}{\delta n} $.  For general $n$ there is no simple relation between the two functionals.  However, when $n$ satisfies the density functional condition
\begin{equation} \frac{\delta F_0}{\delta n({\bf x})}  + [\frac{\delta (F_{Hartree} [n] + F_{xc} [n])}{\delta n({\bf x})} - V_{BO} ] = 0 , \end{equation} then the KS orbitals indeed satisfy
\begin{equation}  (- \nabla^2 + \mu  - \frac{\delta F_0}{\delta n({\bf x})}) \psi_i = \epsilon_i \psi_i . \end{equation}

We conclude that, at the minimum, the KS functional coincides with that obtained by using $F_0 [n]$ rather than the orbital kinetic energies.  The KS equations are equivalent to those obtained by varying the density\footnote{I suspect that this is well known to adepts in DFT, but I could not find a clear explanation of this point in the literature.}.  The KS procedure makes the term $F_0 [n]$ a rather explicit functional of their variational orbitals, while computing $F_0 [n]$ is a nontrivial project, because it's tantamount to solving the single particle Schr\"{o}dinger equation for an arbitrary potential.  KS reserve this work for their numerical solution of the variational equations.  Historically, this was surely motivated by the existence of large computer codes for solving the Hartree equations at the time KS began their work.  

In order to carry out computations in the large $K$ expansion, we need to compute $L[\chi (t, {\bf x}) ]$.  For time independent $\chi$ this is the Legendre transform of $F_0 $.  Thus, for our purposes, working in terms of $n$ rather than KS orbitals, would seem to be the route to follow.

\section{Expansions of the Functional Determinant}

Functional determinants can be computed simply in two different limits, constant fields of arbitrary strength, leading to the derivative expansion, and weak fields of general functional form.  We begin by showing that these two expansions are expansions in a single parameter around $0$ and infinite values.  Consider the Euclidean Schr\"{o}dinger equation
\begin{equation} [\partial_t - \nabla^2 - \mu - g \chi (t, {\bf x}) ] f_n = \lambda_n f_n .\end{equation}
For small $g$ we can solve this by simple perturbation theory. To understand the large $g$ limit we rescale $t \rightarrow t/g$ , $\mu \rightarrow g \mu$, ${\bf x} \rightarrow 1/g {\bf x}$. In terms of the new variables, the equation is
\begin{equation} [\partial_t - \nabla^2 - \mu -  \chi (t g^{-1}, {\bf x} g^{- 1/2}) ] f_n = g^{-1} \lambda_n f_n .\end{equation}  The overall rescaling of the eigenvalues leads to an additive, $\chi$ independent term in the logarithm of the determinant, which we can discard.  It's clear from this formula that the large $g$ limit is equivalent to one in which the fields are slowly varying functions so the derivative expansion, implemented via the heat kernel expansion as in \cite{dietz}, is the appropriate approximation.  The leading term gives the Thomas Fermi approximation to the density functional.

It's certainly possible to compute several terms in the small $g$ expansion fairly easily.  The large $g$ expansion is more complicated but we can certainly get the first non-leading term with relative ease.  My goal is to find an interpolation between these two approximations which will be valid for all $g$, with small error.  There are two different ways to think about such an interpolation.  One can calculate physical quantities like the ground state energy of a particular system as expansions in $g$ and $1/g$ and find interpolating formulae for each quantity individually.  Alternatively, one can try to find an interpolating formula for the entire {\it functional} $L[\chi ]$, and then plug that into the $1/K$ expansion. The second procedure is more universal, and adheres more closely to the original spirit of DFT, but it may be difficult to implement.

\section{1/K expansion of the HEG}

Let us now apply the large spin expansion to the ground state of the HEG. The leading order leads to a translation invariant equation for the background Coulomb photon field $\sigma $
\begin{equation} \nabla^2 \sigma = \frac{1}{T} \frac{\delta L}{\delta \tilde{\sigma} ({\bf x})}  , \end{equation}
where the determinant is calculated with $V_{ext} = \mu$ and a background $\sigma $ independent of Euclidean time.  It is therefore proportional to the Euclidean time interval, $T$, which is taken to infinity.  The tilde over $\sigma$ reminds us that the zero momentum mode is decoupled from the determinant, because of the constraint of neutrality.  The only parameter in the equation is the chemical potential, or equivalently the expectation value of the density operator.

The homogeneous solution of this equation, $\sigma = 0$ gives the correct ground state for high density.  On the other hand, an intuitive argument shows that for low density, a periodic solution, in which there are electron density fluctuations of order $N$ above and below the neutralizing homogeneous positive background charge, is energetically favored.   The scaling argument, which enabled us to scale out the Bohr Rydberg units, shows that the Hartree term in the equation is more important than the non-interacting electron density functional at low density.  Intuitively, the Hartree term gives negative energy from a regular array of positive and negative charge excesses.  One pays a price for this in electron kinetic energy because the band structure in the self consistent periodic potential gives a higher Fermi energy than entirely free electrons.  However, as the length of the lattice spacing goes to infinity, this effect goes away, and the scaling argument shows that it goes away more rapidly than the negative Coulomb energy.

Thus, at sufficiently low density there is a state with negative energy of order $K$, which is the true ground state,  since the order $K$ contribution to the homogeneous ground state energy vanishes.   The quantum phase transition between these two systems is of first order.  The inhomogeneous phase sets in
at a finite wavelength, determined by the critical density.  Gaussian fluctuations around the homogeneous phase are controlled by the Random Phase approximation to the Coulomb photon polarization.  The operator
$$ p^2 + \Pi (p,\omega) , $$ where $\Pi $ is give by a single fermion loop , has no negative modes for any value of the density, so the homogeneous phase is meta-stable in the large $K$ limit.   This is consistent with the finite, $o(K)$, jump in energy between the two phases.

I have not studied the details of the equation closely enough to do more than quote conventional expectations (see the Wikipedia article) for the precise form of the lattice that minimizes the energy at low density.  These are the body centered cubic lattice in three dimensions, the triangular lattice in two, and a uniform lattice in one dimension.  The large $K$ limit gives a controlled context in which the arguments that lead to these expectations, should become rigorous.  

Another point, which one should be able to investigate quite simply in this limit, is whether there is a tricritical point in the density temperature plane, at which the  
transition becomes second order and disappears.  Certainly, we expect that at high enough temperature, only a homogeneous phase exists.  In the large $K$ approximation, the critical quantity to study is the temperature dependent Coulomb photon inverse propagator $p^2 + \Pi (p, \omega, T) $ in the homogeneous phase. As one approaches the critical point along any trajectory in the density-temperature plane, this operator should develop a zero mode and then an unstable mode as one passes through in directions where the crystal phase exists.  It's entirely possible that this computation already exists in the literature, so I will put it off to a future paper until I've done a more thorough literature search.  

Higher order terms in the large $K$ expansion around the homogeneous phase are given in terms of Feynman diagrams.  The propagator in these diagrams is the inverse of $p^2 + \Pi (p, \omega)$, and the vertices are connected correlation functions of the density operator in the free electron gas.  $\Pi$ itself is the two point connected density correlator in the free gas. In a future paper, I intend to compute a few orders in the large K expansion of the ground state energy of the HEG, and compare them to the accurate numerical answers from quantum Monte Carlo.

\section{Conclusions}

I've summarized DFT in a language familiar to quantum field theorists and proposed several strategies for finding analytic approximations for the density functional. I've also suggested that the effective action for the Coulomb photon field might be an even simpler quantity to calculate.  It seems to me that there is a lot more that can be done, and that field theorists should be able to do a really accurate job of calculating at least the connected $n$ point functions of the density operator of the HEG for $n \leq 4$ and the first few terms in the derivative expansion of the effective action, to all orders in the density.

Ignorant of the fundamental results of DFT, I'd always considered the HEG an academic problem of limited interest, but I now realize how central it is to all {\it ab initio} calculations of the properties of matter under the normal circumstances that prevail on earth.  It's worth putting a lot of effort into this problem.  I hope others will agree.

\section{Acknowledgments}
I would like to thank D. Vanderbilt, D. Ceperly, K. Burke, J. Perdew, N. Andrei, P. Coleman, S. Shastry, and especially A. Zangwill, for guidance to the literature on DFT. This work was supported in part by a grant from the Department of Energy.  I would like to thank the Physics Dept. of the Georgia Institute of Technology for its hospitality during the time this paper was written.




\end{document}